\documentstyle[12pt,psfig,twoside]{article}
\begin{document}\hbadness=10000
\pagenumbering{arabic}
\thispagestyle{empty}
\pagestyle{myheadings}\markboth{J. Rafelski, J. Letessier and A. Tounsi}
{QGP Formation and Strange Antibaryons}
\title{QGP formation and strange antibaryons}
\author{$\ $\\
\bf Johann Rafelski$^{1}$, Jean  Letessier$^2$ {\rm and} Ahmed
Tounsi$^2$\\ $\ $\\
$^1$Department of Physics, University of Arizona, Tucson, AZ 85721\\ 
$^2$Laboratoire de Physique Th\'eorique et Hautes Energies\thanks{\em
Unit\'e  associ\'ee au CNRS UA 280.}\\
Universit\'e Paris 7, 2 place Jussieu, F--75251 Cedex 05.\\ }
\date{Published in Phys. Lett B390 (1997) 363}
\maketitle 
\begin{abstract}
{\noindent
We analyze current experimental results and explore, as function of the 
collision energy and stopping in relativistic nuclear collisions,  the
production yields of strange antibaryons, assuming formation of a
deconfined thermal QGP-fireball which undergoes a sudden hadronisation. 
 }
\end{abstract}
We present here a brief account of a study  of the 
hadronic probes of quark-gluon plasma (QGP) involving in particular
strange antibaryons. Hadronic particles can probe the QGP phase provided
that there is rapid final state disintegration --- in this case the
abundances and spectra of hadrons reflect the conditions in the QGP
\cite{gammas,sudden}. 
 
We assume thermalization of initially participating hadronic matter, and
the formation in the collision at currently accessible energies
$\sqrt{s}\le$ 10A GeV of a baryon-rich fireball in the central rapidity
region. The latter hypothesis is inferred from the observed \cite{NA35}
rapidity distribution  of  $\Lambda$ and $\overline{\Lambda}$\,. These
and other results show that the S--Ag/W/Pb and even the S--S  collisions
up to 200A GeV are very different from the ultra-relativistic limit
\cite{Bjo83}, in  which the valence quarks are expected to leave the
central rapidity region. Furthermore, the anticipated \cite{StrRaf} and
experimentally confirmed \cite{NA35,WA85Casc,WA85Om} high production rate
of (multiply) strange antibaryons in A--A reactions, and their central
(in rapidity) spectral distributions are indicating a collective particle
formation mechanism \cite{RD87}: in the QGP reaction picture it is the
ready made high density of (anti) strange quarks which leads under the
rapid hadronisation scenario to high yields of (multiply) strange
particles\cite{StrRaf}.
  
Particle yields from a rapidly dissociating fireball are proportional to
the QGP particle fugacities $\lambda_i$, $i={u,\,d,\,s}$: the fugacity of
hadronic particles is the product of the valence quark fugacities, e.g.
hyperons have the fugacity $\lambda_{\rm Y}=\lambda_{\rm u}\lambda_{\rm
d}\lambda_{\rm s}$. Because of $u$--$d$ symmetry just one light quark
fugacity $\lambda_{\rm q}^2\equiv\lambda_{\rm d}\lambda_{\rm u}$ will be
considered here. One often uses the chemical potentials $\mu_i$:
$\lambda_{i} =e^{\mu_{i}/T}$\,. The chemical potentials for particles and
antiparticles are opposite  to each other, provided that chemical
equilibrium has been established: we speak of {\it relative} chemical
equilibrium when particle abundances are in relative equilibrium with
each other, and of {\it absolute} equilibrium when the total particle
yields are completely filling the available phase space. Calculations 
\cite{sprodQGP,sprodQGPa} show that strangeness will not always fully
saturate the  available phase-space. Therefore, we consider the
associated off-equilibrium parameter $\gamma_{\rm s}$ \cite{gammas}:
\begin{equation}\label{gamth}
\gamma_{\rm s}(t)\equiv {
           \int d^3\!p\,d^3\!x\,n_{\rm s}(\vec p,\vec x;t)\over 
     \int d^3\!p\,d^3\!xn_{\rm s}^{\infty}(\vec p,\vec x)}\ ,
\end{equation}
where $n_{\rm s}^\infty$ is the equilibrium particle density. This
definition presupposes that since the thermal equilibrium is established
within a shorter time scale than the (absolute) chemical equilibrium, the
saturation of the (strangeness) phase space can be described by a
momentum independent factor. The above definition applies in analogous
fashion to light quarks ($\gamma_{\rm q}$) and gluons ($\gamma_{\rm G}$).
It is straightforward to extract from the strange antibaryon experimental
particle yields \cite{analyze} the value of $\gamma_{\rm s}$. The
presence of valance quarks in the fireball helps light quarks and gluons
to equilibrate during the duration of the  nuclear collision, and at
$t_{\rm ch}\simeq 1.5$ fm/c in the CM frame we take $\gamma_{\rm q}\to
1$, $\gamma_{\rm G}\to 1$, but $\gamma_{\rm s}\simeq 1/7$, appropriate
for a 7 times slower strange quark relaxation time \cite{sprodQGPa}. 
 
The specific energy $E/B$ in the fireball is initially only a function of
the CM-energy $E_{\rm CM}$, and the stopping fractions of energy
$\eta_{\rm E}$ and baryon number $\eta_{\rm B}$:
\begin{equation} \label{ECM}
{E\over B}= {\eta_{\rm E}{E_{\rm CM}}\over {\eta_{\rm B}A_{\rm part}}}
\,,\end{equation}
where $A_{\rm part}$ is the number of nucleons participating in the
reaction. When the projectile is smaller than the target, we have assumed
a collision with the geometric target tube of matter. In the case that
the stopping fractions are equal $\eta_{\rm E}\simeq \eta_{\rm B}$ the
resulting specific fireball energies are shown in the heading of the
columns in table \ref{bigtable}: $2.6<E/B<8.8$ GeV. These values
correspond to, in turn: Au--Si and  Au--Au collisions at AGS, possible
future Pb--Pb collisions at SPS with 40A GeV, S--Pb at 200A GeV, and for
the Pb--Pb collisions at 158A GeV, we considered two possible  values of
energy/momentum stopping $\eta=3/4$ and  $\eta=1$\,. 
 
A second condition constraining the initial conditions of the fireball
arises from the balance between the fireball thermal pressure and the
pressure due to kinetic motion:
\begin{equation}
 P_{\rm dyn}=\eta_{\rm p} \rho_{0}{p_{\rm CM}^2}/{E_{\rm CM}}\,.
\end{equation}
Here a fraction $0\le\eta_{\rm p}\le 1$ of the incident CM momentum can
be used by a particle incident on the central  fireball in order to exert
dynamical pressure\cite{init}. In the relativistic limit we have
$\eta_{\rm p}\simeq\eta_{\rm E}\equiv \eta$ since the momentum and energy
of a particle are nearly equal when the velocity $v\simeq c$.
 
Because the QGP phase fireball is initially also strangeness neutral we
have $\lambda_{\rm s}=1$. Therefore, the above two conditions fix the two
open statistical parameters of the fireball $T_{\rm ch}$ and 
$\lambda_{\rm q}$, shown in the columns of the top section of table
\ref{bigtable}. Each column also shows other interesting properties
(number of gluons per baryon, number of light quarks and antiquarks per
baryon, number of anti-strange quarks per baryon, the pressure in the
fireball, baryon density and the entropy per baryon) of the fireball. 
 
At the end of QGP evolution, $t_0\le$ 5 fm/c the strange quarks are near
to absolute  chemical equilibrium abundance and the temperature dropped
from $T_{\rm ch}$ to the value $T_0$ as shown in  the bottom portion of
the table \ref{bigtable}: full chemical equilibrium ($\gamma_s=$ 1) is
here assumed (with exception of the S--W case for which experimental
results imply  $\gamma_s=$ 0.8  \cite{analyze}). During the formation of
the strangeness flavor we allow for fireball evolution,  keeping the
entropy content of gluons and light quarks constant. Conversion of energy 
into strangeness lowers the value of temperature to $T_0$ shown in table.
This temperature is obtained as if the concurrent expansion cooling did
not occur.
\begin{table}[t]
\caption{Properties of different collision fireballs.
\protect\label{bigtable}}
\begin{center} 
\begin{tabular}{|c||c|c|c|c|c|c|} 
\hline
Phase&&\multicolumn{5}{|c|}{\phantom{$\displaystyle\frac{E}{B}$}$E/B$ [GeV]
\phantom{$\displaystyle\frac{E}{B}$}}\\\cline{3-7}
\raisebox{1mm}{space}&\raisebox{2mm}{$<\!s-\bar s\!>=0$}
	&2.6&4.3 & 8.8 & 8.6 & 8.6 \\
occupancy&$\lambda_s\equiv 1$ &$\eta=$ 1 
	&$\eta=$ 1&$\eta\!=\! 0.5$&$\!\eta\!=\! 0.75\!$&$\eta=$ 1 \\
&&Au--Au&Pb--Pb&S--Pb&Pb--Pb&Pb--Pb\\
\hline\hline
&$T_{ch}$ [GeV]&0.212&0.263&0.280&0.304&0.324\\
$\gamma_q=$ 1&$\lambda_q$&4.14&2.36&1.49&1.56&1.61\\
&$n_g/B$&0.56&1.08&2.50&2.24&2.08\\
&$n_q/B$&3.11&3.51&5.16&4.81&4.62\\
$\gamma_g=$ 1&$n_{\bar q}/B$&0.11&0.51&2.16&1.81&1.62\\
&$n_{\bar s}/B$&0.05&0.11&0.25&0.22&0.21\\
&$\!P_{ch}\!$ [GeV/fm$^3$]&0.46&0.76&0.79&1.12&1.46\\
$\gamma_s=$ 0.15&$\rho_{\rm B}$&3.35&3.31&1.80&2.45&3.19\\
&$S/B$&12.3&19.7&41.8&37.4&34.9\\
\hline\hline
&$\gamma_s$&1&1&0.8&1&1\\
$\gamma_q=$ 1&$T_0$ [GeV]&0.184&0.215&0.233&0.239&0.255\\
$\gamma_g=$ 1&$\lambda_q$&4.14&2.36&1.49&1.56&1.61\\
&$n_{\bar s}/B$&0.34&0.68&1.27&1.43&1.33\\
$\gamma_s=$ 0.8&$\!P_0\!$ [GeV/fm$^3$]&0.30&0.41&0.47&0.54&0.71\\
or&$\rho_{\rm B}$&2.17&1.80&1.05&1.19&1.56\\
$\gamma_s=$ 1&$S/B$&14.5&24.0&49.5&46.5&43.4\\
\hline
\end{tabular} 
\end{center} 
\end{table}

The temperature $T_0$ is closely related to the observed inverse slope
$T_\bot$ of the transverse mass spectrum fitted using the thermal
spectral form $dN/m_{\bot}^{3/2}dm_\bot\propto \exp(-m_\bot/T_\bot)$.
Namely, the presence of some transparency (longitudinal flow) causes the
central rapidity region $\Delta y=1$ to feature in the $m_\bot$ spectra
the thermal shape normally seen after a sum over full rapidity range,
here $\Delta y=3$. Furthermore, we  have in transverse direction either
directly the emission of particles and hence $T_{\bot}\simeq T_0$, or
there is collective radial flow in the QGP-hot matter, in which fraction 
of the thermal energy is converted into the  flow energy.  When  the
final state  particles emerge from the expending volume and without
re-equilibration, their spectral shape comprises the QGP temperature,
blue-shifted by the flow velocity. This Doppler shift effect restores the
freeze-out temperature to the high initial values\cite{ULI93}, while
absence of re-equilibration eliminates the reheating or cooling effects. 
For the S--W/Pb collisions at 8.8 GeV we have $T_0=233$ MeV, which result
agrees very well with the experimental value $T_\bot=232\pm5$
\cite{WA85Casc}. Similarly $\lambda_{\rm q}=1.49$ is in agreement with
the results of our previous data analysis   \cite{analyze}. This
agreement is a consequence of the choice $\alpha_s=0.6$ in the QGP-EoS
and stopping $\eta=1/2$.
 
In  Fig.\,\ref{fig1S95} we show, as function of the specific energy
content $E/B$, the behavior of temperature $T_0$,  the light quark
fugacity $\lambda_{\rm q}$ and entropy per baryon $S/B$ at full chemical
equilibration in the QGP fireball. The range of the possible values as
function of $\eta$ is indicated by showing  results, for $\eta=1$ (solid
line), 1/2 (dot-dashed line) and 1/4 (dashed line). We note that, in
qualitative terms, the drop in temperature with decreasing energy and
stopping is intuitively as expected, and the value of $\lambda_{\rm q}$
is relatively insensitive to the stopping power. We note the (rapid) rise
of specific entropy with $E/B$ which should lead to a noticeable excess
in the final particle abundances at CERN-SPS energies \cite{entropy}.

The experimental bars show for high (8.8 GeV) energy the result of WA85
data analysis  \cite{analyze}, and for low energy (2.6 GeV) are taken 
from the analysis of the BNL-AGS data   \cite{BNL-AGSthermal}. Note  that
in this case we had found $\lambda_{\rm s}=1.7$ and not $\lambda_{\rm
s}=1$ as would be needed for the QGP interpretation  at this low energy:
Aside of the light quark fugacity there is also the strange quark
fugacity not shown in Fig.\,\ref{fig1S95}, since in a strangeness neutral
QGP fireball $\lambda_{\rm s}=1$\,, independent of the prevailing
temperature and baryon density. This happens in general when both $s$ and
$\bar s$ quarks have the same phase-space size, which is the case e.g.
when they exist unbound. On the other hand, in confined forms of hadronic
matter at finite baryon density there is a strong constraint between the
two fugacities $\lambda_{\rm q}$, and $\lambda_{\rm s}$ arising from the
requirement of strangeness conservation \cite{analyze}. These non-trivial
relations between the parameters characterizing the final state particle
abundances \cite{gammas,analyze} are in general difficult to
satisfy, and thus lead to particle abundances which  differ considerably
between different final states. It is of importance to note that the two
different collision systems analyzed at 200A GeV  (S--W and S--S)
\cite{NA35,WA85Casc}  lead to \cite{analyze} $\lambda_{\rm s}\simeq 1$ at
different $T_\bot$.  Our explanation of this result is that the particle 
source is a rapidly dissociating, deconfined fireball.
 
We now explore as function of energy the production of (strange) baryons 
and antibaryons. The ratios of (strange) antibaryons to strange baryons
{\it of same  particle type\/}: $R_{\rm N}=\bar p/p$,
$R_\Lambda=\overline{\Lambda}/ \Lambda$,  $R_\Xi=\overline{\Xi}/\Xi$ and
$R_\Omega=\overline{\Omega}/ \Omega$, are \cite{gammas,analyze}  simple
functions of the quark fugacities. The behavior of these ratios is shown
in  Fig.\,\ref{eqratios}{\bf a} as function of energy. It is obtained
using the results for $\lambda_{\rm q}$  shown in  Fig.\,\ref{fig1S95},
and taking the QGP value $\lambda_{\rm s}=1$. We have to remember that
$R_\Omega=\lambda_{\rm s}^{-6}=1$, but since  some re-equilibration  is
to be expected towards the HG behavior $\lambda_{\rm s}>1$, we expect
$\lambda_{\rm s}=1+\epsilon$, with $\epsilon$ small, and thus for this
ratio $R_\Omega=1-6\epsilon<1$. A further non negligible correction which
has been discussed in Ref.\, \cite{analyze} is due to the isospin
asymmetry.
 
In order to assess, as function of collision energy, the magnitude of the
strange antibaryon yields per particle multiplicity formed in the
collision, we need to establish the excitation function of the individual
particle (antibaryon) yields.  Considerable uncertainty is arising from
the off-equilibrium nature of the hadronisation process, which in
particular makes it hard to estimate how the different heavy particle
resonances are populated. Some of these uncertainties are eliminated when
we normalize the yields in Fig.\,\ref{eqratios}{\bf b} at an energy,
which we take here to be the value $E/B=2.6$ GeV (we assume freeze-out
temperature $T=150$ MeV, $\gamma_{\rm s}=1,\, \eta_{\rm p}=1/2$ and
absence of any re-equilibration after particle production). These yields
increase with energy, as would be also expected in cascading hadron
interactions, but contrary to such a picture in our approach the rise of
more strange antibaryon yield is less pronounced.

We finally consider in  Fig.\,\ref{sratios} the particle ratios involving
particles with differing strangeness content, and which are thus
sensitive to the strangeness equilibration. The error bars show the key
experimental results obtained at $\sqrt{s}=8.6$A GeV, compared to our
theoretical  results obtained with same phase space cuts on the range of
$p_\bot$ as in the experiments and presented here as function of energy.
The $\overline{\Lambda}/\bar p\simeq 0.8\pm 0.25$  ratio of the NA35
collaboration \cite{NA35pbar} was obtained for the S--Au system at 200A
GeV for full phase space; WA85 precise value 
$\overline{\Xi^-}/\overline{\Lambda}=0.21\pm0.02$ for $p_\bot>1.2$ GeV,
which result determines our choice $\gamma_{\rm s}=0.70 $ and  to a
lesser degree also $\eta_{\rm p}=1/2$; and WA85 \cite{WA85Om} 
$(\Omega+\overline{\Omega})/(\Xi^-+\overline{\Xi^-})=0.8\pm0.4$ for
$p_\bot>1.6$ GeV.  The fact that the  two ratio $\overline{\Lambda}/\bar
p$ (NA35) and $(\Omega+\overline{\Omega})/(\Xi^-+\overline{\Xi^-})$
(WA85) are satisfactorily  explained, provides a very  nice confirmation
of  the consistency of the thermal fireball model. We also draw attention 
to the remarkable behavior of the  $\overline{\Xi^-}/\overline{\Lambda}$ 
ratio, which rises rapidly as the energy decreases.
 
The large $\overline{\Xi}/ \overline{\Lambda}$ ratios in our 
QGP-fireball reaction picture are found even at relatively small
energies, as is seen in Fig.\,\ref{sratios}. This is in contrast to
microscopic models  --- near to  $\overline{\Xi}$ production threshold in
$p$--$p$ interaction this ratio is exceedingly small. This lets us expect
that as the energy rises towards the QGP formation threshold, there ought
to be a considerable discontinuity in the relative $\overline{\Xi}/
\overline{\Lambda}$ yield as function of collision energy. This provides
for an interesting possibility to  identify the energy at which
collective QGP based production of strange antibaryons is first
encountered. 
 
At this threshold energy we should observe the onset of the other
specific features of the deconfined hadronic phase: strange phase space
saturation ($\gamma_{\rm s}\to 1$)\,, the associated strange particle
production  enhancement, pattern of strange antibaryon flow showing
$\lambda_{\rm s}=1$\,, and entropy enhancement (particle multiplicity
enhancement). If these features are indeed found, we can safely conclude
that we have discovered a novel hadronic phase comprising deconfinement
of $s$ and $\bar s$ quarks, rapid strangeness production, collective
hadron production, high entropy content, as is expected of a rapidly
dissociating quark-gluon plasma fireball.

\vspace{0.5cm}
\subsection*{Acknowledgment}
 J.R. acknowledges partial support by  DOE, grant
		DE-FG03-95ER40937 \,. 

\begin{figure}[p]
\caption{\protect\small 
Temperature $T_0$, light quark fugacity $\lambda_{\rm q}$
and entropy per baryon $S/B$ at the time $t_0\simeq 5$ fm/$c$ of absolute 
chemical equilibration, as function of the QGP-fireball energy content 
$E/B$; stopping $\eta=1$ (solid line), 1/2 (dot-dashed line) and 1/4
(dashed line). See text for comparison with analysis results.
\protect\label{fig1S95}
}
\caption{\protect\small 
{\bf a)} Antibaryon to baryon abundance ratios as function of energy per 
baryon $E/B$: $R_{\rm N}=\bar p/p$ (solid line),
$R_\Lambda=\overline{\Lambda}/\Lambda$ (long-dashed line), 
$R_\Xi=\overline{\Xi}/\Xi$ (short-dashed line)
 and $R_\Omega=\overline{\Omega}/\Omega$  (dotted line).
{\bf b)} Relative antibaryon yields as function of $E/B$: $\overline{p}$
(solid line), $\overline{\Lambda}$  (long-dashed line) $\overline{\Xi^-}$
(short-dashed line) and $\overline{\Omega}$ (dotted line), all normalized
to their respective yields at $E/B=2.6$ GeV\,.  \protect\label{eqratios}
}
\caption{\protect\small 
Strange antibaryon ratios for S--W/Pb collisions as function 
of $E/B$:
$\overline{\Lambda}/\overline{p}$ (full phase space),
$\overline{\Xi^-}/\overline{\Lambda}$ for $p_\bot>1.2$ GeV and
$(\overline{\Omega}+\Omega)/(\overline{\Xi^-}+\Xi^-)$ for
$p_\bot>1.6$ GeV; experimental results 
shown are from experiments NA35, WA85, see text for details.
 \protect\label{sratios}}
\end{figure}
\newpage
\begin{figure}[p]
\vspace*{1cm}
\centerline{\hspace*{-1.5cm}
\psfig{width=13cm,figure=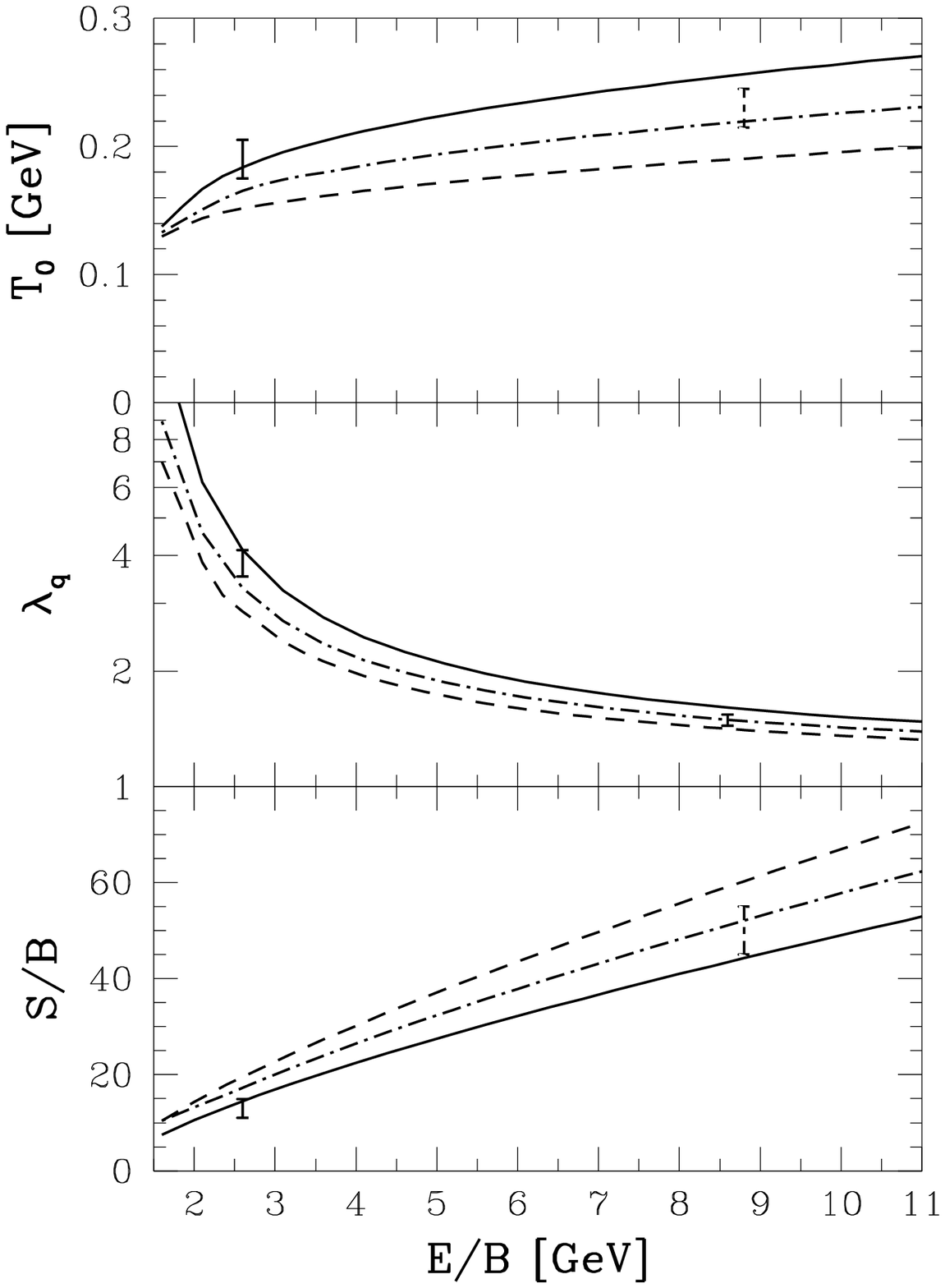}}
\vspace*{-0cm}
\centerline{\bf FIGURE 1}
\end{figure}
\newpage
\begin{figure}[p]
\vspace*{-4cm}
\centerline{\hspace*{-0.6cm}
\psfig{width=16cm,figure=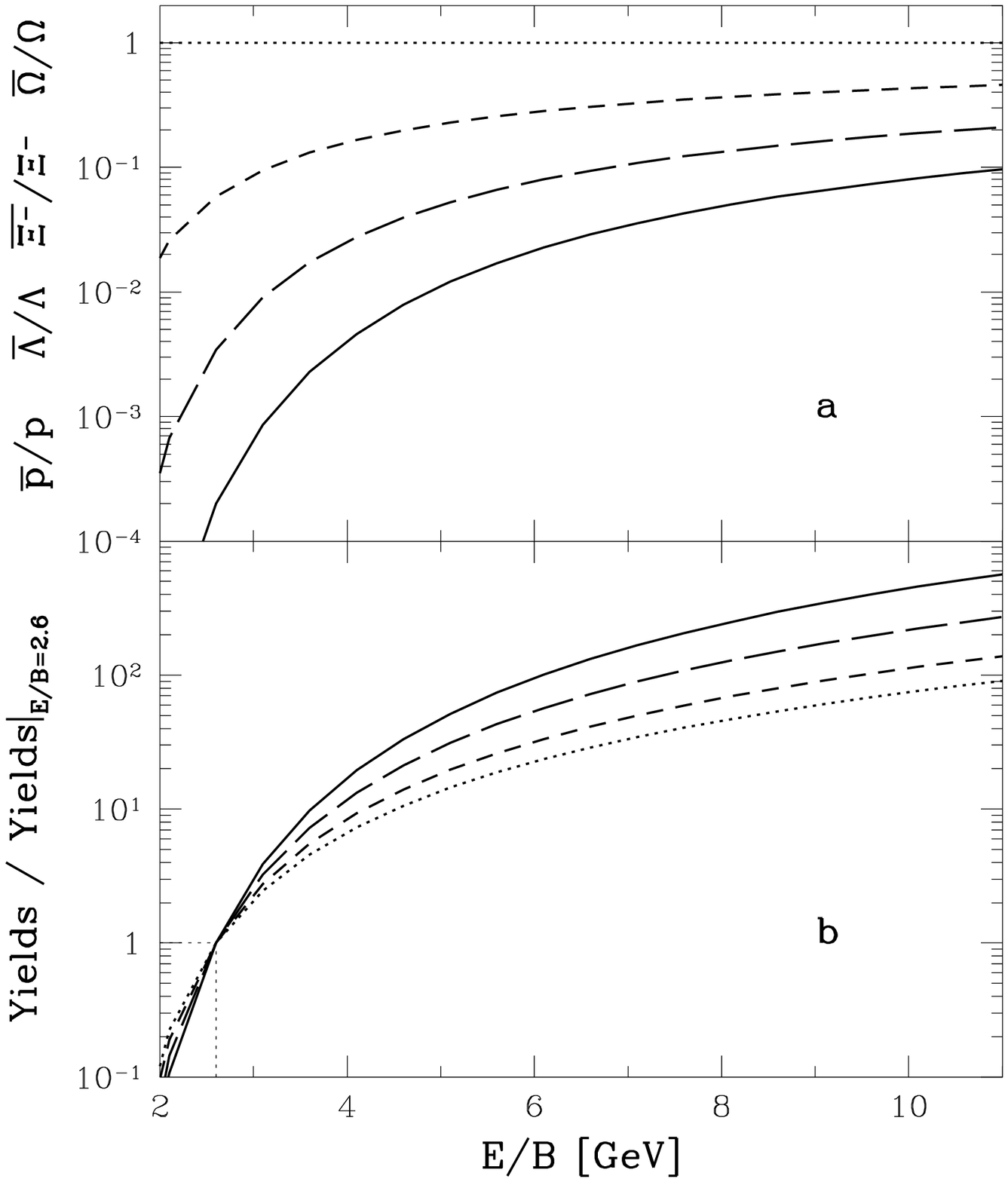}}
\vspace*{0.6cm}
\centerline{\bf FIGURE 2}
\end{figure}
\newpage
\begin{figure}[p]
\vspace*{3cm}
\centerline{\hspace*{2cm}
\psfig{width=17cm,figure=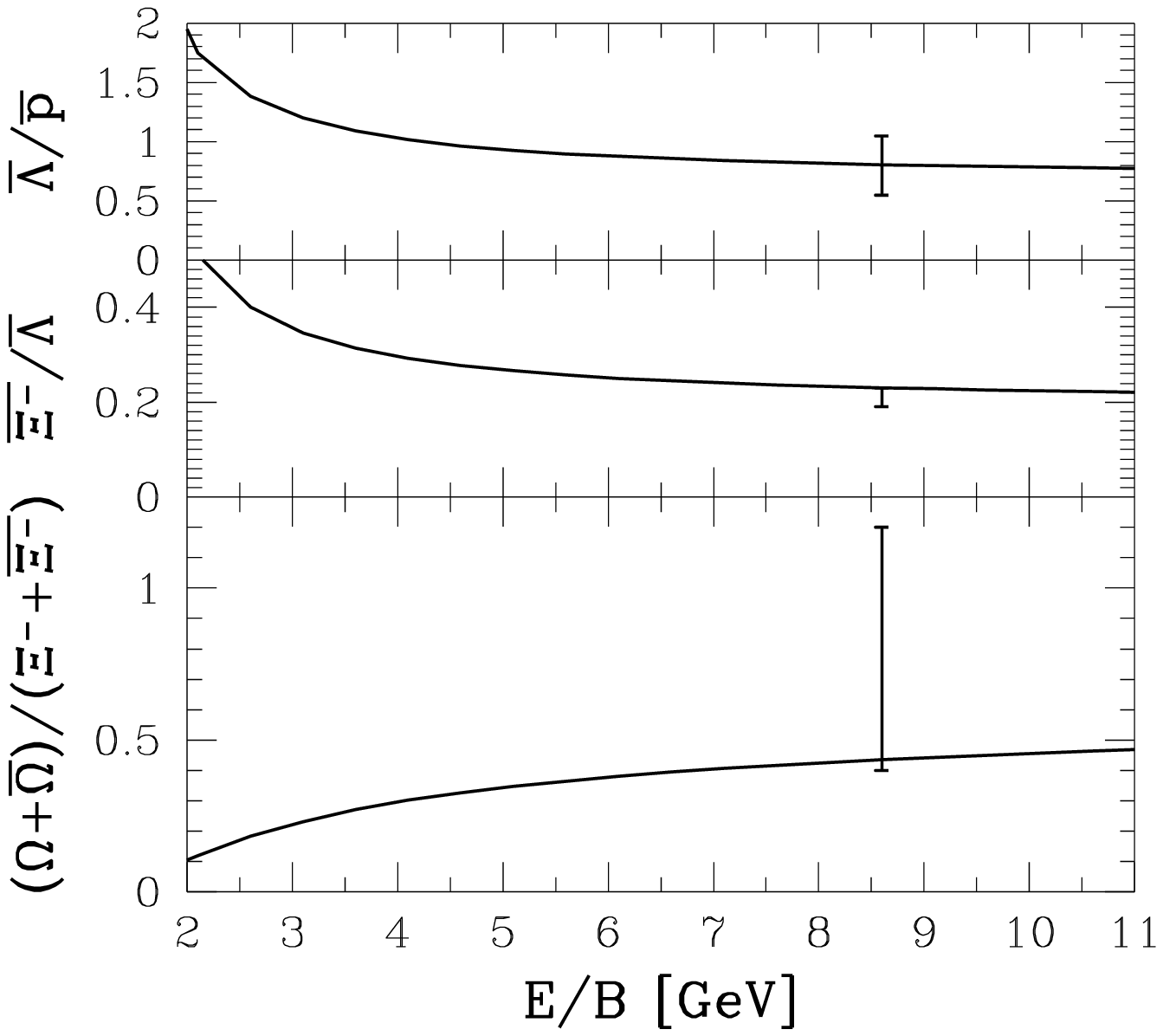}}
\vspace*{-2cm}
\centerline{\bf FIGURE 3
\vspace*{-2cm}
}
\end{figure}



\begin{thebibliography}{99}
 
\bibitem{gammas} J. Rafelski, {Phys. Lett.} B {\bf 262},  333 (1991);
{Nucl. Phys.} A {\bf 544}, 279c (1992).
 
\bibitem{sudden} L. P. Csernai and I. N. Mishustin, Phys. Rev. Lett. {\bf
74}, 5005 (1995).
 
\bibitem{NA35} T. Alber et al. (NA35 collab.) 
Z. Phys. C {\bf 64}, 195 (1994).
 
\bibitem{Bjo83} J.D. Bjorken, {Phys. Rev.} D {\bf 27}, 140 (1983).  
 
\bibitem{StrRaf} J. Rafelski and R. Hagedorn, {\it Statistical Mechanics
of Quarks and Hadrons}, North-Holland, Amsterdam 1981, p. 253, edited by
H. Satz ; J. Rafelski, Phys. Rep. C {\bf 88}, 331 (1982).
 
\bibitem{WA85Casc} S. Abatzis et al. (WA94 collab.)
Phys. Lett. B {\bf 354}, 178 (1995); D. Evans {et~al.} (WA85 collab.),
{Nucl. Phys.} A {\bf 566}, 225c (1994).
  
\bibitem{WA85Om}
S. Abatzis {et al.} (WA85 collab.), 
{Phys. Lett.} B {\bf 347}, 158 (1995); 
{Phys. Lett.} B {\bf 316}, 615 (1993).  
 
\bibitem{RD87}
J. Rafelski and M. Danos, {\it Phys. Lett.} B {\bf 192}, 432 (1987); 
{Phys. Rev.} D {\bf 27}, 671 (1983).
 
\bibitem{sprodQGP}J. Rafelski and B. M\" uller, Phys. Rev. Lett.  {\bf
48}, 1066 (1982); {\bf 56}, 2334E (1986).
 
\bibitem{sprodQGPa}N. Bili\'c, J. Cleymans, I. Dadi\'c and D. Hislop,
Phys. Rev. C {\bf 53?},\ldots (1995) and references therein.  
 
\bibitem{analyze} 
J. Letessier, A. Tounsi, U. Heinz, J. Sollfrank and J. Rafelski, 
{Phys.\ Rev.} D {\bf 51}, 3408 (1995);
J. Sollfrank, M. Ga\'zdzicki, U. Heinz and J. Rafelski,  {Z. Physik} C
{\bf 61}, (1994);
J. Letessier, J. Rafelski and  A. Tounsi,  {Phys. Lett.} B {\bf 321}, 394
(1994).
 
\bibitem{eb} 
J. Letessier, J. Rafelski and  A. Tounsi,  {Phys. Lett.} B
{\bf 323}, 393 (1994).

\bibitem{init}  J. Letessier, J. Rafelski and A. Tounsi,
Phys. Lett. B {\bf 333}, 484 (1994).
 
\bibitem{ULI93}  
E. Schnedermann, J. Sollfrank and U. Heinz,  {\it Particle Production in
Highly Excited Matter}, NATO Physics series Vol. B {\bf 303}, Plenum
Press, New York, 1993, p. 175, edited by H.H. Gutbrod and J. Rafelski.
 
\bibitem{entropy}
J. Rafelski, J. Letessier and A. Tounsi, {\it XXVI International
Conference on High Energy Physics}, Dallas, 
Texas, 1992, AIP-Conference Proceedings No 272, p. 983, 
edited by J.R. Sanford;
J. Letessier, A. Tounsi, U. Heinz, J. Sollfrank and J. Rafelski,  {Phys.
Rev. Lett.} {\bf 70}, 3530 (1993);
M. Ga\'zdzicki, Z. Phys. C {\bf 66}, 659, (1995).
 
\bibitem{BNL-AGSthermal} 
J. Rafelski and M. Danos {Phys. Rev.}, C {\bf 50}, 1684 (1994);\\   J.
Letessier, J. Rafelski and A. Tounsi, {Phys. Lett.} B {\bf 328}, 499
(1994);
P. Braun-Munzinger and J. Stachel, in {\it Hot Hadronic Matter},
p451, NATO-ASI Series B: Physics Vol. 346, Plenum Press, New York 1995,
edited by J. Letessier, H.H. Gutbrod and J. Rafelski.
 
\bibitem{NA35pbar} T. Alber et al. (NA35 collab.), preprint
IKF-HENPG/5-95, submitted to Phys. Lett. {B}; J. G\"unther et al. (NA35
collab.), to appear in proceedings of QM'95, Monterey, January 1995,
edited by A. Poskanzer et al..
 
 

\end{thebibliography}
\end{document}